\documentclass{aip-cp}

\usepackage[numbers]{natbib}
\usepackage{rotating}
\usepackage{graphicx}

\begin{document}

\title{Pulsar Magnetospheres and Pulsar Winds}

\author[aff1,aff2]{Vasily Beskin}
\eaddress{beskin@lpi.ru}

\affil[aff1]{P.N.Lebedev Physical Institute, Leninsky prosp., 53, Moscow, 119991, Russia} 
\affil[aff2]{Moscow Institute of Physics and Technology, Institutsky per., 9, Dolgoprudny,
141700, Russia}

\maketitle

\begin{abstract}
Surprisingly, the chronology of nearly 50 years of the pulsar magnetosphere and pulsar
wind research is quite similar to the history of our civilization. Using this analogy, I 
have tried to outline the main results obtained in this field. In addition to my talk, 
the possibility of particle acceleration due to different processes in the pulsar 
magnetosphere is discussed in more detail. 
\end{abstract}

\section{INTRODUCTION}

After 50 years of radio pulsar research, our understanding remains elusive. In 
spite of significant progress which had been achieved, no consensus regarding 
such key questions as the nature of the coherent radio emission or the conversion 
of electromagnetic energy to particle energy yet exists. Here I have tried to 
outline the main results that have been obtained in this field. Due to space 
limitations, I do not discuss at all very important processes of the interaction 
of the pulsar wind with supernova remnant. For the same reason only names, not 
full references, are given.

\section{PULSAR CHRONOLOGY}

\subsection{Ancient world (before 1967)}

Long before the discovery of radio pulsars by J.Bell and A.Hewish in 1967 it was 
already clear that neutron stars (mass $M \sim M_{\odot}$, radius $R \sim 10$ 
km) are to exist. They were predicted by W.Baade and F.Zwicky as early as in 
30th{\footnote{L.Landau has published his famous paper several months before the 
neutron was discovered.}}. Moreover, F.Pacini predicted their rotation periods 
$P \sim 1$ s and magnetic fields $B_{0} \sim 10^{12}$ G. But noone could imagine 
that solitary neutron stars are to be very active cosmic sources, mainly in radio 
band. For this reason, specific observations were not carried out, and radio pulsars 
were serendipitously discovered within another observational program. Remember that 
the possibility to detect neutron stars in binary systems was clearly formulated, 
and X-ray pulsars were immediately discovered when the first space X-ray observatory 
was launched. 

\subsection{Hellas (1968--1973)}

It was wonderful epoch of simple images which allowed to intuitively understand 
the nature. Magnetized sphere rotating in a vacuum was a sufficient model to 
clarify the main properties of radio pulsars. So, it became clear that it is a 
neutron star rotation that gives rise to the extremely stable sequence of radio 
pulses, and that the kinetic energy of rotation provides the reservoir of energy. 
Simultaneously, the key idea that the energy losses are connected with electrodynamic 
processes was formulated. Up to now the textbook relation for vacuum magneto-dipole 
energy losses 
\begin{equation}
W_{\rm tot}^{({\rm V})} = -I_{\rm r} \Omega \dot\Omega 
=  \frac{1}{6} \, \frac{B_0^2 \Omega^4 R^6}{c^3} \sin^2 \chi 
\approx  10^{32} B_{12}^2 \, P^{-4} \, \, {\rm erg}/{\rm s}
\label{Eqn01}
\end{equation}
(where here and below $B_{12} = B_{0}/10^{12}$ G, ${\dot P}_{-15} = {\dot P}_{-15}/10^{-15}$,
and period $P$ is in seconds) is used for evaluation of the energy losses of radio pulsars. 
Remember that the moment of truth was just connected with relation (\ref{Eqn01}) because after 
discovery of Crab pulsar ($P = 33$ ms, ${\dot P}_{-15} \approx 420$) two already known numbers, 
namely, energy budget $5 \, 10^{38}$ erg/s which is necessary to inject into Crab Nebula to 
explain its optical emission by synchrotron radiation and dynamical life time  
$\tau_{\rm D} = P/2{\dot P} \sim 1000$ years corresponding to historical supernova 1054AD, 
found their natural explanation.

By the way, this toy model helped also to make the first step in understanding radio 
pulsars as sources of cosmic rays. Indeed, for millisecond pulsars ($P \sim 1$ ms) and 
high enough magnetic fields $B_{0} \sim 10^{13}$ G the potential difference between 
the pole and equator of a rotating sphere $\Delta V \approx (\Omega R/c)B_{0}R$ does 
reach $10^{19}$ eV, i.e., almost the maximum energy which is observed in cosmic rays. 
The same estimate was obtained by J.Gunn \& J.Ostriker for particle acceleration in 
the electromagnetic wave outgoing from such young rotating magnetized neutron star. 

However, in a few years it became clear that magnetized sphere rotating in 
a vacuum is an oversimplification of the reality. In 1971, P.Sturrock recognized 
the importance of one-photon pair production in superstrong magnetic field 
$\gamma +(B) \rightarrow e^{+} + e^{-} + (B)$. Otherwise, the chain of processes 
is:

1. primary particle acceleration by the longitudinal electric field $E_{\parallel}$ 
induced by neutron star rotation,

2. emission of $\gamma$-quanta by primary particle moving along curved magnetic field line
with characteristic frequency $\omega \sim (c/R_{\rm cur})\gamma_{\rm e}^3$ (the so-called 
'curvature radiation'),

3. photon propagation in the curved magnetic field and its subsequent decay into 
the secondary $e^{+}e^{-}$ pair; synchrotron photons, which are produced when 
secondary pairs go into zeroth Landau level, are also important,

4. secondary particles acceleration and emission of curvature photons giving rise to new 
generations of particles.

\noindent
This process is to create enough secondary electron-positron pairs to screen efficiently 
accelerating electric field $E_{\parallel}$. Of course, this also implies that radio pulsars 
are to be sources of positrons, but with much smaller energy. 

Thus, radio pulsars are to have magnetosphere filled with plasma. This implies that to zeroth
order the longitudinal electric field can be considered to be exactly screened ($E_{\parallel} = 0$).
The occurrence of the longitudinal electric field in some region immediately leads to an abrupt 
plasma acceleration and to the explosive  generation of secondary particles. Due to the screening, 
the plasma is to corotate rigidly with the neutron star, as it is observed in the Earth and Jovian 
magnetospheres. Just this property gives us the key to understand the pulsed activity of radio 
pulsars. 

\begin{figure}[h]
\centerline{\includegraphics[width=430pt]{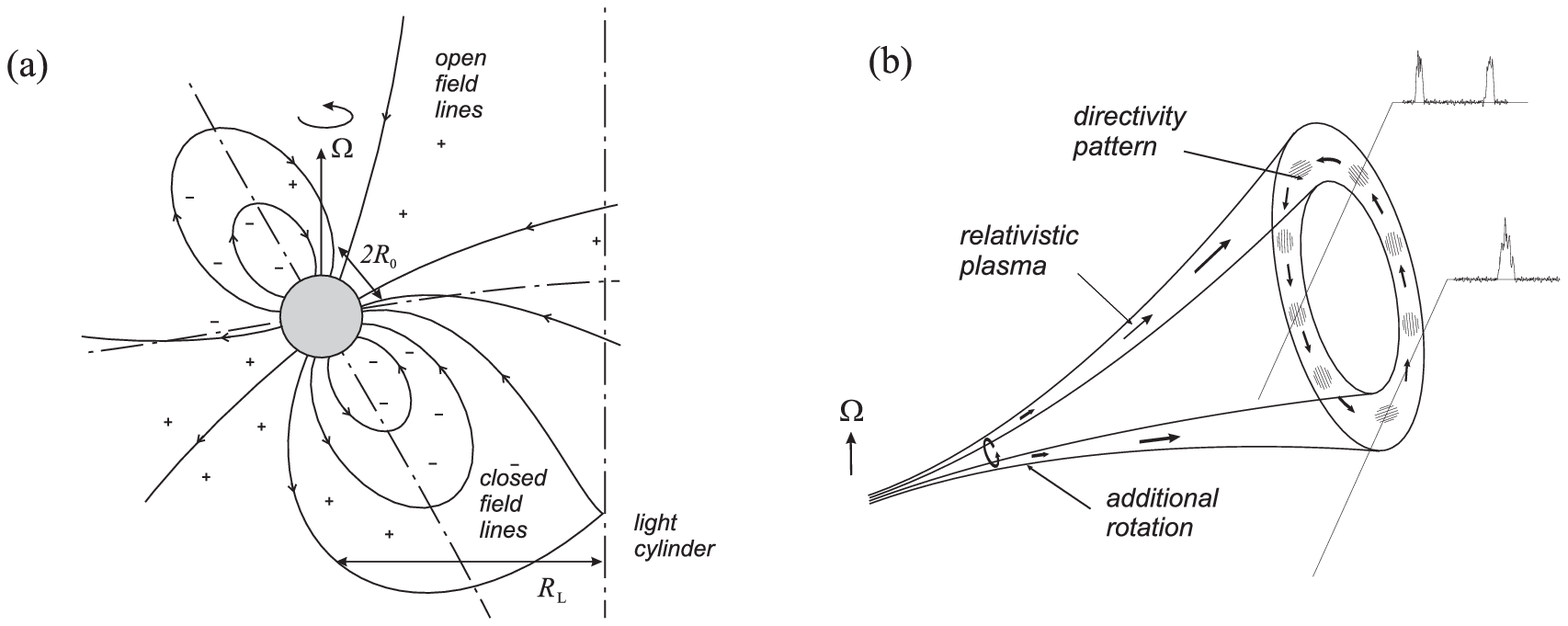}}
\caption{
The main elements of the pulsar magnetosphere (a) and geometrical 'hollow cone' model of the
directivity pattern (b).} 
\label{fig01}
\end{figure}

Indeed, it is clear that the rigid corotation becomes impossible at large distances from the 
rotation axis $r_{\perp} > R_{\rm L}$, where $R_{\rm L} = c/\Omega$ is so-called light cylinder 
radius (see Fig.\ref{fig01}a). For ordinary pulsars $R_{\rm L} \sim$ $10^{9}$--$10^{10}$ cm, 
i.e., the light cylinder locates at the distances several thousand times larger than the neutron 
star radius. Further, we can estimate the polar cap size $R_{0} \approx (\Omega r/c)^{1/2}R$, 
i.e., the region at the magnetic pole of a neutron star, from which the magnetic field lines 
extend beyond the light cylinder. For ordinary radio pulsars the polar cap size is only several 
hundreds of meters. Importance of the polar cap region connects with the possibility for 
charged particles to escape from the neutron star magnetosphere. Indeed, as particles can 
move only along magnetic field lines (not only because of the smallness of the Larmor radius 
in comparison with another characteristic scales, but also due to very short synchrotron 
life time resulting in the drop to zero Landau level), two types of magnetic field lines 
appear in the magnetosphere. Open field lines start at the polar cap, smoothly intersect 
the light cylinder and extend to infinity. Closed field lines originate far from the magnetic 
axis and closed within the light cylinder. The plasma located on the closed magnetic 
lines is trapped, whereas the plasma filling the open magnetic lines can escape from the 
neutron star magnetosphere. 

This simple model allowed V.Radhakrishnan \& D.Cooke (and later L.Oster \& W.Sieber) 
to formulate so-called 'hollow cone' model of the pulsars radio emission, which perfectly 
accounted for their basic geometric properties. Indeed, the secondary plasma generation 
is to be suppressed in the very vicinity of the magnetic pole  where, first, the intensity 
of the curvature radiation is low due to approx. rectilinear magnetic field lines  and, 
second, the photons emitted by relativistic particles propagate at small angles to the 
magnetic field. Therefore, as shown in Fig.\ref{fig01}b, in the central regions of the 
open magnetic field lines a decrease in the secondary plasma density should be expected.
If we  now make a rather reasonable assumption that the radio emission is directly connected
with the outflowing plasma density, there must be a decrease in the radio emission intensity 
in the center of the directivity pattern. Therefore, without going into details (actually, 
the mean profiles have a rather complex structure), we should expect a single (one-hump) 
mean profile in pulsars in which the line of sight intersects the directivity pattern far 
from its center and the double (two-hump) profile for the central passage. This is exactly 
what is observed in reality.

In the same years three main parameters determining key physical 
electrodynamical processes were formulated 
\begin{equation}
\rho_{\rm GJ} = -\frac{{\bf \Omega \cdot B}}{2 \pi c}, \qquad 
\lambda = \frac{n_{\rm e}}{n_{\rm GJ}}, \qquad 
\sigma_{\rm M} = \frac{1}{4 \lambda} \, \frac{e B_{0}\Omega^{2} R^{3}}{m_{\rm e}c^4}.
 \label{Eqn02} 
\end{equation}
The first one is the charge density which is necessary to screen longitudinal electric 
field. This fundamental quantity introduced by P.Goldreich \& P.Julian in 1969 
determines also the characteristic number density $n_{\rm GJ} = |\rho_{\rm GJ}|/|e|$ 
(about $10^{12}$ cm$^{-3}$ near the star surface) and, what is much more important, 
characteristic current density $j_{\rm GJ} = c \rho_{\rm GJ}$. As it will be shown 
below, it is the electric current circulating in the pulsar magnetosphere that plays 
the main role in our play. The second parameter is the multiplicity $\lambda$ which 
shows to what extent the number density of the secondary plasma exceeds the critical 
number density $n_{\rm GJ}$. And the third one is so-called Michel magnetization parameter 
$\sigma_{\rm M}$. It corresponds to maximum bulk Lorentz-factor $\Gamma_{\rm max}$ 
which can be achieved if all the energy losses $W_{\rm tot}$ (\ref{fig01}) transferred 
into bulk hydrodynamical particle flow  ${\dot N} m_{\rm e}c^2 \Gamma$, where 
${\dot N} = \lambda \pi R_{0}^2 n_{\rm GJ}c$ is the electron-positron ejection rate. 
Now the notation $\sigma = \sigma_{\rm M}/\Gamma = W_{\rm em}/W_{\rm part}$ is used, 
so we add subscript 'M' for clearness.

Thus, in first several years after discovery of radio pulsars the answers to key questions 
were obtained. The only problem was in recognizing how does the pulsar magnetosphere work...
Indeed, it was necessary to understand how the energy is transfeered from rotating neutron 
star to infinity, what is the energy spectrum of outflowing particles, and, certainly, 
what is the mechanism of the radio emission (which is to be coherent due to extremely 
high brightness tempetature up to $T_{\rm b} \sim 10^{28}$ K). Up to now the answers
to most these questions remain unknown...

\subsection{Rome (1973--1983)}

Rome was the epoch when first strict laws were formulated. It concerned all main subjects 
including pair creation region, pulsar magnetosphere, and pulsar wind. At first, two detailed 
models of the pair creation in the vicinity of the neutron star surface were formulated. 
This process becomes possible due to continuous escape of particles along the open field lines. 
As a result, the domain (a 'gap') with longitudinal electric field forms in the vicinity of the 
magnetic poles, its height being determined by the secondary plasma generation mechanism. 
As thought that time, a greater part of secondary particles is generated above the 
acceleration region, where the longitudinal electric field is rather small, so that the 
secondary plasma can freely escape from the neutron star magnetosphere. 

First detailed model was developed in 1975 by M.Ruderman \& P.Sutherland, and also by 
V.Eidman group. In their models the particle ejection from the neutron star surface plays 
no role as the electron work function was thought to be high enough. Later in 1978-1983, 
where more accurate theoretical consideration has shown that electron work function can be 
small, alternative model postulating free ejection of particles from the neutron star 
surface was developed by J.Arons group. It is the last model that was considered as the 
most adequate one within next three decades in spite of some intrinsic problems (e.g., in 
its first version it gave pair creation on the north half of the polar cap only).

The importance of these models was in determination of multiplicity $\lambda$ and the spectrum 
of secondary plasma. Starting from the first results obtained by J.Daugherty \& A.Harding in 
1982 it became clear that multiplicity cannot exceed $10^{4}$--$10^{5}$. This implies that the 
number density near the star surface $\lambda n_{\rm GJ} \sim 10^{16}$ cm$^{-3}$ is too low for 
detection 511 keV annihilation line. Accordingly, the Michel magnetization parameter 
$\sigma_{\rm M}$ is to be as large as $10^{3}$--$10^{4}$ for ordinary pulsars and can reach 
$10^{6}$ for fast young pulsars (Crab, Vela) only. As a result, the electron-positron 
injection rate ${\dot N}$ and the total injection $N = \int {\dot N}{\rm d}t$ per 
individual radio pulsar during all its life time can be presented as
\begin{equation}
{\dot N} \approx \lambda \, 
\frac{B_{0}\Omega^2 R^3}{ce} \sim 3 \cdot 10^{34} \, P^{-2} B_{12} \, {\rm s}^{-1}, 
\qquad 
N \approx \lambda \, \frac{Mc^2}{e B_{0}R} \sim 10^{50} \, B_{12}^{-1},
 \label{Eqn03} 
\end{equation}
the last value depending very weak on initial period. As the total number of neutron 
stars in the Galaxy does not exceed $10^{9}$, one can conclude that radio pulsars cannot 
be the main sources of Galactic positrons. But pulsars may be relevant for PAMELA excess 
of energetic positrons in the energy domain 1--100 GeV, what is widely discussed now. 

\begin{figure}[h]
\centerline{\includegraphics[width=400pt]{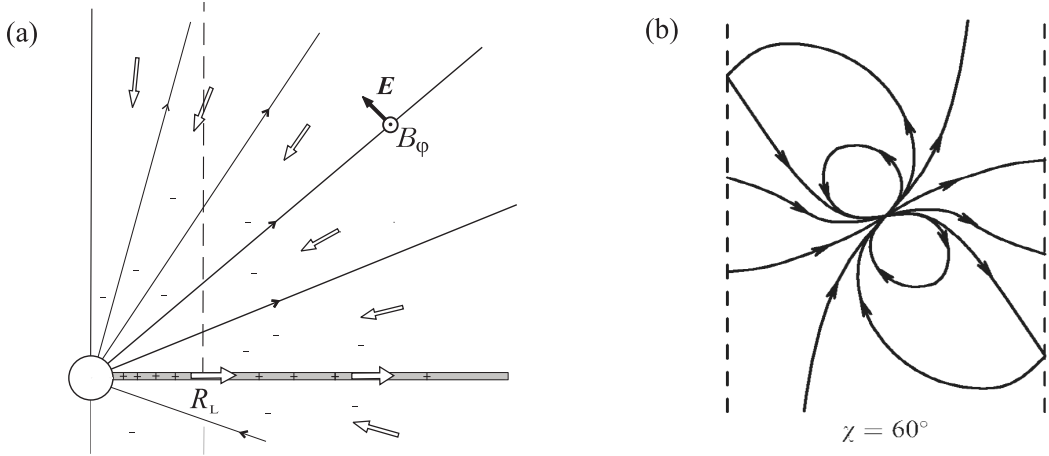}}
\caption{Michel split monopole solution (a) and BGI inclined solution with zero
longitudinal electric current (b).} 
\label{fig02}
\end{figure}

Important results were obtained in the theory of the pulsar magnetosphere and 
pulsar wind. First, L.Mestel, F.Michel, I.Okamoto and many other researchers 
formulated axisymmetric force-free 'pulsar equation' which became the main 
instrument in theoretical consideration. In patricular, one of its analytical 
solutions (which could be found only for a few model profiles of the longitudinal 
current $j_{\parallel}$) have shown that for GJ longitudinal current 
$j_{\parallel} = j_{\rm GJ}$ one can construct quasi-spherical wind solution 
in which electric field is smaller than magnetic one up to infinity 
\begin{equation}
B_{r} = B_{0} \frac{R^2}{r^2}, \quad
B_{\varphi} = E_{\theta} = 
- B_{0} \, \frac{\Omega R^2}{c r} \, \sin\theta, \quad
j_{r} = \rho_{\rm GJ} c = 
- \frac{\Omega B_{0}}{2 \pi} \, \frac{R^2}{r^2}\, \cos\theta, \quad
S_{r} = 
\frac{B_{0}^2\Omega^2 R^4}{4 \pi c r^{2}} \sin^2\theta. 
\label{Eqn04} 
\end{equation}
As shown in Fig.\ref{fig02}a, in this solution longitudinal electric currents (contour 
arrows) generate toroidal magnetic field $B_{\varphi}$ which together with inductive electric 
field $E_{\theta}$ forms radial Poynting flux $S_{r}$ taking the energy away from a neutron 
star. Thus, the very possibility of the MHD outflow up to infinity was demonstrated. 
In such axisymmetric (i.e., time-independent) force-free solution all energy losses 
connect with the Poynting flux. But contrary to magneto-dipole wave, it is realised
at zero frequency. Of course, the presence of the equatorial sheet separating outgoing in
ingoing magnetic fluxes seemed rather artificial. But as we will see, this solution
helped a lot in constructing self-consistent solution.

Besides, our team (V.Beskin, A.Gurevich \& Ya.Istomin, hereafter BGI) have solved 
analytically the inclined version of the 'pulsar equation' (see Fig.\ref{fig02}b). 
It was found that for zero longitudinal electric current circulating in the neutron star 
magnetosphere the energy losses vanish for any inclination angle. This effect (confirmed 
later by L.Mestel group) results from full screening of the magneto-dipole radiation 
by magnetospheric plasma. This implies that the braking of the neutron star rotation 
results fully from impact of the torque ${\bf K}$ due to longitudinal currents circulating 
in the pulsar magnetosphere. It is convenient to describe two components of the torque 
${\bf K}$ parallel and perpendicular to the magnetic dipole ${\bf m}$ by dimensionless 
current in the polar cap zone $i = j_{\parallel}/j_{\rm GJ}$ separating it into symmetric 
and antisymmetric contributions, $i_{\rm s}$ and $i_{\rm a}$, depending upon whether the 
direction of the current is the same in the north and south parts of the polar cap, or 
opposite. As one can easily check,
$K_{\parallel} \propto i_{\rm s}$, and  $K_{\perp} \propto i_{\rm a}$. Here and 
below we apply normalization to the `local' Goldreich-Julian current density, 
$j_{\rm GJ} = |{\bf \Omega}\cdot{\bf B}|/2\pi$ (with scalar product). In particular, 
the direct action of the Amp\`ere force on the star 
${\bf K} = \int [{\bf r} \times [{\bf J}_{\rm s} \times {\bf B}]/c] \, {\rm d}{\bf S}$
by surface currents ${\bf J}_{\rm s}$ which close the longitudinal electric currents 
circulating in the pulsar magnetosphere can be written as
\begin{eqnarray}
K_{\parallel}^{\rm sur}  \approx  -\frac{B_{0}^{2}\Omega^{3}R^{6}}{c^{3}} 
i_{\rm s},\label{16'} \quad 
K_{\perp}^{\rm sur}  \approx - \frac{B_{0}^{2}\Omega^{3}R^{6}}{c^{3}}
\left(\frac{\Omega R}{c}\right)i_{\rm a}.
\label{17'}
\end{eqnarray}
In particular, for `local' GJ current ($i_{\rm s} \approx i_{\rm a} \approx 1$) 
relations (\ref{17'}) imply 
$K_{\perp}^{\rm sur} \approx \left(\Omega R/c\right) K_{\parallel}^{\rm sur}$.
Below we also assume (as was not done up to now) that the additional contribution for 
$K_{\perp}$ can give the magnetosphere itself, more precisely, the mismatch between 
magneto-dipole radiation from magnetized star and radiation generated by charges 
filling the pulsar magnetosphere (they exactly compensate themselves for 
$j_{\parallel} = 0$). Here we write down $K_{\perp}^{\rm mag}$ in general form as
\begin{eqnarray}
K_{\perp}^{\rm mag}  =  - 
A \, \frac{B_{0}^{2}\Omega^{3}R^{6}}{c^{3}} \, i_{\rm a}
\label{17''}
\end{eqnarray}
trying to evaluate the dimensionless constant $A$ later from the results of numerical 
simulations. 

Returning now to time evolution of the angular velocity $\Omega$ and the inclination 
angle $\chi$, one can write down
\begin{eqnarray}
I_{\rm r} \, \dot{\Omega} =  
K_{\parallel}^{\rm A} + (K_{\perp}^{\rm A}-K_{\parallel}^{\rm A})\sin^2\chi,
\quad I_{\rm r} \, \Omega \, {\dot\chi} =  
(K_{\perp}^{\rm A}-K_{\parallel}^{\rm A})\sin\chi\cos\chi,
\label{Eqn05}
\end{eqnarray}
where $I_{r} \sim M R^2$ is the neutron star momentum of inertia and we put 
$K_{\parallel} = K_{\parallel}^{\rm A} \cos\chi$ and $K_{\perp} = K_{\perp}^{\rm A} \sin\chi$. 
As both expressions contain the factor $(K_{\perp}^{\rm A}-K_{\parallel}^{\rm A})$, 
inclination angle $\chi$ is to evolve to $90^{\circ}$ (counter-alignment) 
if the total energy losses decrease for larger inclination angles, and to $0^{\circ}$ 
(alignment) if they increase with inclination angle. E.g., for local GJ longitudinal current
($i_{\rm s}^{A} \approx i_{\rm a}^{A} \approx 1$), as was supposed by BGI, we have 
counter-alignment evolution:
\begin{equation}
W_{\rm tot}^{({\rm BGI})} = 
i_{\rm a}^{\rm A}(\Omega) \, \frac{f_{\star}^{2}(\chi)}{4} \,
\frac{B_{0}^2\Omega^4 R^6}{c^3} \, \cos^2\chi,
\quad
{\dot \chi}^{({\rm BGI})} =
i_{\rm a}^{\rm A}(\Omega) \, \frac{f_{\star}^{2}(\chi)}{4 I_{\rm r}} \,
 \frac{B_{0}^{2}\Omega^{2}R^{6}}{c^{2}} \sin\chi \, \cos\chi,
\quad
1.59 < f_{\star}(\chi) < 1.96.
\label{Eqn07}
\end{equation}

Thus, during these years, it was able to formulate the basic laws of the pulsar 
magnetosphere. Moreover, several analytical solutions were constructed and even 
some quantitative predictions were formulated. It seemed that very soon we will 
understand the main properties of the magnetosphere of radio pulsars. Unfortunately, 
the problem was much more complicated. As a result, due to the lack of significant 
progress in the solution of nonlinear equations describing pulsar magnetosphere 
(which could not be solved analytically, and numerical methods have not yet been 
sufficiently developed) led to the abrupt decrease of a number of scientists working 
in this area. The dark ages came.

\subsection{Dark ages (1983--1999)}

Those were indeed dark ages, especially in the theory of the pulsar magnetosphere. At 
first glance, no results were obtained in that 15 years at all. But slowly, step by 
step, our understanding of the magnetospheric processes became more and more clear.  
First, important results were obtained in the theory of magnetized winds. Remember that 
force-free (i.e., massless) approach told nothing about the energy of outflowing particles. 
Only more general MHD theory, which was elaborated very intensively that time, demonstrated 
that in the quasi-spherical magnetized wind there is no effective particle acceleration. 
As was shown by A.Tomimatsu in 1994, at large distances (more exactly, outside the fast 
magnetosonic surface $r \gg r_{\rm F}$, where $r_{\rm F} \sim \sigma_{\rm M}^{1/3} \, R_{\rm L}$) 
particle energy cannot exceed $\sigma_{\rm M}^{1/3} \, m_{\rm e} c^2$, i.e.,
Poyting-to-particle energy flux ratio $\sigma = W_{\rm em}/W_{\rm part}$
is to be large: $\sigma \sim \sigma_{\rm M}^{2/3} \gg 1$. Simultaneously, it was recognized 
that the londitudinal electric current $j_{\parallel}$, as the accretion rate in Bondi 
accretion, is not a free parameter, but is to be fixed by crirical condition at the fast 
magnetosonic surface. In relativistic case it must be close to GJ current $j_{\rm GJ}$. 

Another step ahead was connected with recognizing the importance of General 
Relativity (GR) in dynamics of the pair creation process. This effect results 
from additional term appearing in the expression for the GJ charge density 
$\rho_{\rm GJ} \approx -(\Omega - \omega)B/2\pi \,c$ connected with Lense-Thirring 
angular velocity $\omega$ . In spite of its smallness, appropriate derivative could 
be large enough. As was shown by V.Beskin in 1990 and by A.Muslimov \& A.Tsygan 
in 1992, in Arons model the pair creation process becomes possible within all polar 
cap region just resulting from GR effects.  

Besides, in the theory of the pulsar wind several very important results were 
obtained. At first, F.Coroniti and F.Michel have recognized that in the absence
of the magneto-dipole wave the wind outflow from inclined magnetized neutron star is 
to contain 'striped' current sheet separating ingoing and outgoing magnetic fluxes. 
Later C.Kennel \& F.Coroniti analysing interaction of the pulsar wind with Crab 
nebula have found that at large distances compared with the dimension of the nebula the 
magnetization of the pulsar wind is to be very low: $\sigma \sim 10^{-2}$. It was 
in direct disagreement with theory prediction mentioned above. This '$\sigma$-problem',
i.e., the impossibility to accelerate plasma effective enough in the quasi-radial pulsar 
wind, remains one of the main problem of the theory of the pulsar magnetopshere. At any 
way, within MHD approximation it cannot be solved.

\begin{figure}[h]
\centerline{\includegraphics[width=400pt]{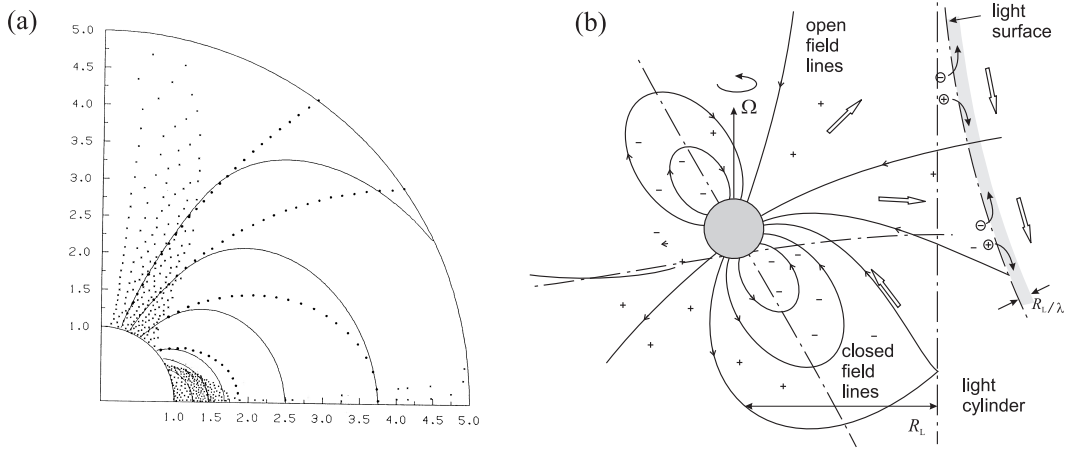}}
\caption{Two 'crazy' ideas: Michel \& Krause-Polstorff 'disk-dome' structure of the pulsars 
magnetosphere (a) and BGI magnetospheric structure with light surface where very effective 
particle acceleration and current closure take place (b).} 
\label{fig03}
\end{figure} 

As often happens in difficult times, several 'crazy' ideas were suggested (see 
Fig.\ref{fig03}). First, F.Michel together with J.Krause-Polstorff in 1984-1985 
has considered so-called 'disk-dome' structure of the pulsar magnetosphere in which 
positive and negative charges are trapped in different regions of the magnetosphere, 
which are separated by vacuum region{\footnote{Qualitatively this structure was 
considered by Yu.Rylov and E.Jackson in 1970-ies.}}. At first glance, it was a step 
back in comparison with previuos works. In particular, it was unclear whether this 
configuration is stable in the presence of pair formation or not. But, as we will see, 
this structure was reproduced in recent numerical simulations. 

Besides, BGI group considered the case where longitudinal current $j_{\parallel}$ is 
small enough so that MHD flow cannot be realized up to infinity. In this case, as is 
also shown in Fig.\ref{fig03}, the magnetosphere has a 'natural boundary' --- the 
light surface, where electri\ field becomes equal to magnetic one, and where very 
effective bulk particle acceleration up to $\Gamma \approx \sigma_{\rm M}$ takes 
place (solving the $\sigma$-problem!); as we shall see, this assumption also 
subsequently found its indirect confirmation. This structure inevitably appears for 
the inclined rotator and for Arons model, which postulated exactly local GJ longitudinal 
current density $j_{\parallel} = c \rho_{\rm GJ} \approx \Omega B \cos\chi/2\pi$, 
i.e., the current which is small enough to support MHD outflow to infinity. Indeed, 
as was shown above, in MHD wind the toroidal magnetic field at the light cylinder is 
to be equal to electric one. But for local GJ electric current (and for inclined rotator) 
the total current is too small to generate large enough toroidal magnetic field. 
Thus, within BGI model the low-$\sigma$ pulsar wind is formed already in the vicinity
of the light cylinder.

\subsection{Renaissance (1999--2006)}

\begin{figure}[h]
\centerline{\includegraphics[width=400pt]{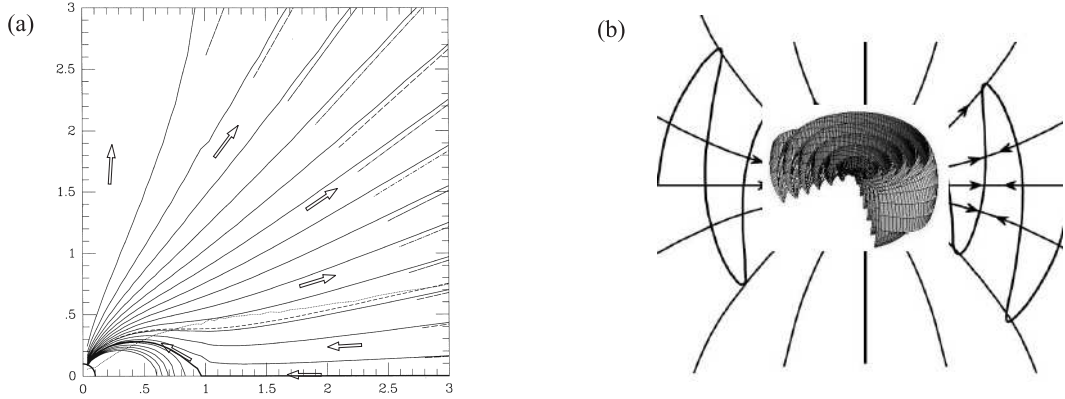}}
\caption{ CKF axisymmetric force-free pulsar magnetosphere; contour arrows show 
electric currents (a) and Bogovalov 'incline split monopole' analytical solution (b).}
\label{fig04}
\end{figure}

This epoch (started with two papers published in 1999) was the time of rapid 
breakthrough in understanding the beauty and the truth. In the first one I.Contopoulos, 
D.Kazanas \& Ch.Fendt (CKF) found numerically (by the iterative procedure) the 
solution of the axisymmetric 'pulsar equation'. Similarly to Michel split-monopole 
solution, at large distances $r > R_{\rm L}$ it contained the equatorial current sheet, 
but near the stellar surface it corresponded to dipolar, not monopolar magnetic field 
(see Fig.\ref{fig04}a). In a few years this structure was reproduced in many 
papers{\footnote{Among 15 autors of these papers 7 are Russian-speaking persons...}} 
rebooting the interest to the theory of the pulsar magnetosphere. Or course, these 
solutions concerned axisymmetric structure which gave us no information about real 
time-dependent magnetospheres of radio pulsars. 

Simultaneously, S.Bogovalov found analytical force-free solution for 'inclined 
split monopole' pulsar wind
\begin{equation}
B_{r} = B_{0} \frac{R^2}{r^2} \, {\rm Sign}(\Phi), \,
B_{\varphi} = E_{\theta} 
= - B_{0} \, \frac{\Omega R^{2}}{c r} 
\, \sin\theta  \, {\rm Sign}(\Phi), \,
\Phi = \cos\theta \cos\chi - \sin\theta \sin\chi 
\cos\left[\varphi - \Omega \left(t - r/c\right)\right],
\label{Eqn08}
\end{equation}
where the condition $\Phi = 0$ just determines the shape of the striped current sheet
(see Fig.\ref{fig04}b). On the other hand, near the star the magnetic field had 
monopolar structure. In this solution within the cones \mbox{$\theta < \pi/2 - \chi$}, 
$\pi - \theta < \pi/2 - \chi$ around the rotation axis the electromagnetic fields 
are time-stationary and coinside with axisymmetric Michel solution, which was already 
shown in Fig.\ref{fig02}a. On the other hand, in the equatorial region all components 
of electromagnetic field change signs at the instant when the current sheet intersects 
given point; between these intersection times the fields remain time-independent. 
It is not superfluous to mention that this solution contained no magneto-dipole 
wave. It is necessary to stress as well that the condition $\Phi = 0$ for the shape of 
the current sheet actually has the kinematic origin. In particular, it is to be true 
for any solution with radial poloidal field lines and asymptotic behavior 
$E_{\theta}(\theta) = B_{\varphi}(\theta)$ with arbitrary $\theta$-dependence as well; 
such time-indepenent asymptotic solution of the 'pulsar equation' was analytically 
obtained by R.Inhragam already in 1973. So, it is not surprising that later the shape 
of the striped current sheet was reproduced in all numerical simulations.

Thus, undeniable progress has been made during these years. At first, it was in 
recognizing the extremely important role of the current sheet in the dynamics of 
the pulsar wind. It is not surprising that analysis of physical processes inside 
the current sheet became the mainstream in the pulsar wind theory. In particular, 
J.Kirk \& Yu.Lyubarsky have mentioned the role of magnetic reconnection, which 
also gave start to numerous investigations. Another key point was connected with the 
determination of 'universal solution' which, as was already stressed, was reproduced 
in many papers. But the most important subject was in changing the very sight on the 
londitudinal current circulating in the pulsar magnetosphere. Longitudinal current 
$j_{\parallel}^{({\rm us)}}$ given by 'universal solution', not by the pair creation 
mechanism was decleared now as the correct value corresponding to the real configuration. 
In other words, it was postulated that the solution with zero longitudinal electric 
field (both force-free and MHD approaches deny that) must satisfy not only the 
electric charge condition $\rho_{\rm e} = \rho_{\rm GJ}$, but also the current conition 
$j_{\parallel} = j_{\parallel}^{({\rm us)}}$. For this reason, no restriction on the 
longitudinal electric current was postulated within all force-free and MHD simulations; 
in the last case plasma was freely injected into domains, where during the simulation 
the plasma number density becomes low enough.

On the other hand, some points were not clarified. At first, 'universal solution' 
gave the profile of longitudinal current density $j_{\parallel}$ which was in 
direct disagreement with Arons model postulating, as was 
specially stressed above, local GJ current density. Simultaneously, it contradicted 
'split monopole' profile as well. In particular, as shown in Fig.\ref{fig04}a, 
'universal solution' predicted volume return current. Remember that common point 
of view in first papers devoted to pulsar magnetosphere was in introducing return 
electric current only along the separatrix separating open and closed magnetic field 
lines. Indeed, all pair creation models postulated the same sign of the accelerating 
potential difference through the polar cap for non-orthogonal rotators and, accordingly, 
the same direction of the longitudinal current. Appearence of the volume return current 
through the polar cap area was rather strange as it was in a direct disagreement with 
the pair creation models. Though this difficulty has not been spoken aloud, first 
attempts were made to resolve this contradiction already at that time. In particular, 
S.Shibata and later A.Beloborodov demonstrated that on field lines with 
small enough longitudinal current $j_{\parallel} < j_{\rm GJ}$, particle acceleration 
and the pair creation in the polar cap process is to be supressed.

\subsection{Industrial revolution (2006--2013)}

\begin{figure}[h]
\centerline{\includegraphics[width=400pt]{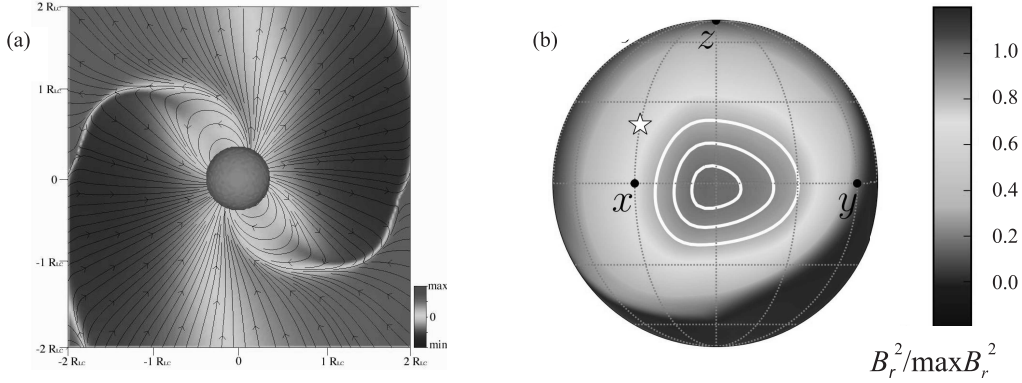}}
\caption{Inclined ($\chi = 60^{\circ}$) Spitkovsky force-free solution (a) and MHD
solution demonstrating concentration of magnetic field lines at $r = 6 \, R_{\rm L}$ 
in the direction shifted in azimuth by about $30^{\circ}$ to magnetic axis (contour star) 
(b).} 
\label{fig05}
\end{figure}

Industrial revolution in pulsar physics was connected with the possibility to produce 3D 
time-dependent simulations; as any revolution, it succeeded in resolving accumulated 
intrinsic problems by purely technical means. This epoch started with a paper published 
in 2006 in which A.Spitkovsky numerically solved force-free equations for inclined 
rotator. It was the first 3D solution for rotating inclined magnetic dipole with 
the wind extending up to infinity. As shown in Fig.\ref{fig05}a, in spite of small 
computation box, this solution confirmed the existence of the striped current sheet.

Industrial revolution is impossible in one separate country. After this paper, 
not only force-free, but also MHD simulations for inclined rotator were produced 
in many scientific centers{\footnote{Mainly in Princeton (A.Spitkovsky, J.Li, 
A.Tchekhovskoy), and also by I.Contopoulos and S.Komissarov groups.}}. As a result, 
'universal inclined solution' was found. In particular, the original result obtained 
by A.Spitkovsky for total energy losses
\begin{equation}
W_{\rm tot}^{({\rm MHD})} \approx 
\frac{1}{4} \, \frac{B_{0}^2\Omega^4 R^6}{c^3} \, (1 + \sin^2\chi),
\quad
{\dot \chi}^{({\rm MHD})} \approx  
-\frac{1}{4I_{\rm r}} \frac{B_{0}^{2}\Omega^{2}R^{6}}{c^{2}} \sin\chi \, \cos\chi
\label{Eqn09}
\end{equation}
was confirmed. As we see, 'universal solution' gives the increase of total energy 
losses with inclination angle $\chi$, and, according to Eqn. (\ref{Eqn05}), its 
aligned evolution (the second expression confirmed this point was obtained later). 

On the other hand, it was shown that 'universal solution' differs drastically from 
the Michel-Bogovalov split-monopole one (\ref{Eqn08}). In particular, for large enough 
inclination  angles $\chi > 30^{\circ}$ the radial magnetic field is compressed to 
equatorial plane, so that approximately
\begin{equation}
<B_{r}>_{\varphi} \approx B_{0} \frac{R^2}{r^2} \, \sin\theta \, {\rm Sign}(\Phi), 
\quad
<B_{\varphi}>_{\varphi} = <E_{\theta}>_{\varphi} 
\approx - B_{0} \, \frac{\Omega R^2}{cr} 
\, \sin^2\theta  \, {\rm Sign}(\Phi), 
\quad
<S_{r}>_{\varphi} \approx 
 \frac{B_{0}^2\Omega^2 R^4}{4 \pi c r^2} \sin^4\theta,
\label{Eqn10}
\end{equation}
while Michel-Bogovalov solution (\ref{Eqn08}) corresponds to homogeneous radial 
magnetic field $B_{r} = B_{0}(R^2/r^2) \, {\rm Sign}(\Phi)$. As was already stressed, 
radial asymptotic solutions with arbitrary $\theta$-dependenses were known since 
70-ties. But it was surprising that the results of 3D time-dependent MHD simulation 
(\ref{Eqn10}) are in good agreement with very simple force-free solution predicting, 
e.g., the general connection $S_{r}(\theta) \propto B_{r}^2(\theta) \, \sin^2\theta$, 
which is just fulfilled. Besides, it turned out that appropriate dimensionless current 
$i_{\rm a}^{\rm A} \sim (\Omega R/c)^{-1/2}$ is insufficient to explain MHD energy 
losses (\ref{Eqn09}) only by the Amp\'ere force torque $K^{\rm sur}$ (\ref{17'}) due 
to polar cap surface currents. Evidently, they can be connected with magnetospheric 
losses (\ref{17''}) if we assume $A \sim (\Omega R/c)^{1/2}$. It is important that 
for such a value $A \ll 1$ on can neglect the contribution of the magnetospheric torque 
$K_{\perp}^{\rm mag}$ (\ref{17''}) within BGI model (\ref{Eqn07}), as was actually done. 

One can stress here that in Eqn. (\ref{Eqn10}) $\varphi$-averaged values are presented. 
In reality, as  shown in Fig.\ref{fig05}b, there is visible concentration of magnetic 
field lines (as well as the Poynting flux) in direction shifted in azimuth by about 
$30^{\circ}$ to magnetic axis (contour star); in other words, in the equatorial region 
between two parts of the current sheet the flow is $\varphi$-dependent. In this point 
there is another very important difference with Michel-Bogovalov solution in which, 
as is clear from Eqn. (\ref{Eqn08}), the electromagnetic fields are not time- and 
$\varphi$-depenent outside the striped current sheet.

At the same time, new sight on the universal current structure allowed A.Timokhin 
(and later A.Timokhin \& J.Arons) to consider in more detail the pair creation 
region in the vicinity of the magnetic poles. Now longitudinal current 
$j_{\parallel}^{({\rm us)}}$ was the input parameter in their consideration. It was 
demonstrated that pair creation is indeed possible for large enough longitudinal 
currents $j_{\parallel} > j_{\rm GJ}$ and, as was already mentioned, not possible 
for smaller ones. Moreover, it was shown that pair creation is possible for volume 
return current as well. It becomes possible due to essential time-dependence of 
the pair creation process. Unfortunately, these were 1D simulations which could 
potentially loose some important features. In particular, it was impossible to 
include into consideration the time-dependence of the toroidal magnetic field. 
Nevertheless, this work was very important in understanding the very possibility 
to produce longitudinal current which differs drastically from local GJ current. 
And one of the main result was associated with an actual return to RS model in which 
particle injection from the star surface does not play a determining role. 

To summarize, one can conclude that industrial revolution helped us to make 
the next step in understanding the structure of the pulsar magnetosphere. 
It is very important that all these results were confirmed by several groups 
independently. Nevertheless, some key problems were not solved. And the main 
one concerned the support of the necessary current $j_{\parallel}^{({\rm us)}}$, 
which was determined by 'universal solution',  by pair creation processes in 
the magnetosphere. 

\subsection{Modern time (2014--)}

Formally, it was only the next step in numerical simulation associated with 3D
particle-in-cell (PIC) code. But actually it was a qualitative step ahead as 
only kinetic treatment, unlike force-free and MHD approaches, could self-consistently 
include into consideration the pair creation process, i.e., the acceleration 
regions with nonzero parallel electric field $E_{\parallel}$ and subsequent particle 
injection into the computation domain. In other words, PIC approach, at least in 
principle, allows to produce ab-initio simulations. Of course, this epoch is in the 
very begining, so not all results presented below were confirmed by independent 
consideration. Besides, in spite of (natural for this stage) childhood diseases, new 
possibilities connected with PIC approach have already given a number of very interesting 
results{\footnote{Again, about a half of researchers working now in pulsar PIC simulation
are Russian-speaking persons...}}.

\begin{figure}[h]
\centerline{\includegraphics[width=400pt]{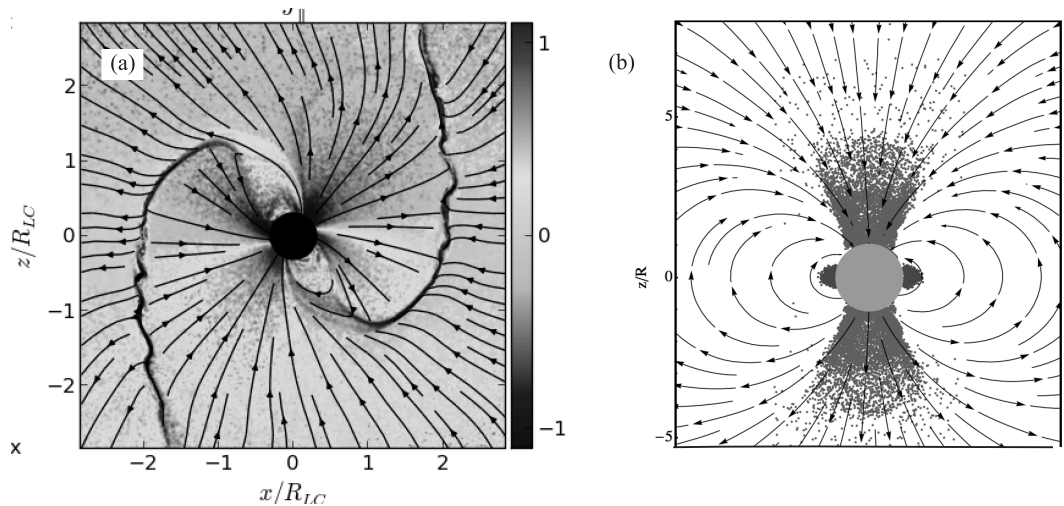}}
\caption{First Princeton PIC simulations: solution for incline rotator (a) and
for axisymmetric case without including GR effects into consideration (b).
} 
\label{fig06}
\end{figure}

At first, as shown in Fig.\ref{fig06}a, A.Philippov, A.Spitkovsky \& B.Cerutti  
have reproduced MHD structure contaning, in particular, striped current sheet. 
These simulation also provided extra confirmation of the existence of the 'universal 
solution'. Second, the possibility of the effective particle acceleration just outside 
the light cylinder up to $\Gamma \approx \sigma_{\rm M}$ was demonstrated as well. This 
process became possible due to the appearence of a domain with high enough electric
field $E > B$ inside the equatorial current sheet. In
addition, reconnection electric field even in the axisymmetric case helps particles to 
drift towards the current sheet. As a result, up to 30\% of the electromagnetic energy 
flux dissipates in the vicinity of the light cylinder $r < 5\, R_{\rm L}$ transferring 
electromagnetic energy flux into particle acceleration. At the moment it is rather 
difficult to say whether this process solves the $\sigma$-problem (it is definitely not 
at 5 light cylinders), but, certainly, some step ahead was done. At any way, such an 
effective particle acceleration definitely helps us to explain high energy radiation 
detected by Fermi observatory. 
 
On the other hand, in some sense PIC simulations have put more questions than give 
answers. Indeed, already the first results obtained by A.Philippov \& A.Spitkovsky 
and by A.Chen \& A.Beloborodov in 2014 for axisymmetric magnetosphere have suddenly shown 
the absence of pair creation in polar region near magnetic poles. In contrast, as one 
can see in Fig.\ref{fig06}b, free ejection from  the neutron star surface (which was 
postulated in their work) resulted in the formation of the 'disk-dome' structure which 
was already discussed above. Later it was shown that pair creation in the bulk of the 
polar cap is absent in the inclined magnetosphere for small enough inclinations 
$\chi < 30^{\circ}$. This effect was connected with a little bit smaller longitudinal 
current $j_{\parallel}^{({\rm us})}$ in comparison with $j_{\rm GJ}$; as was already 
stressed, in this case no particle acceleration and pair creation takes place. 
Fortunately, this problem was soon resolved. It turned out that GR effects (which, as 
was already mentioned, changes the GJ charge density) 
flips the inequality to $j_{\parallel}^{({\rm us})} > j_{\rm GJ}$ allowing thereby the 
pair creation process. As is shown in Fig.\ref{fig06}a, in this case PIC simulation 
reproduced in general the structure obtained within MHD approach.  

The second, more serious problem concerned the formation of the return current. At the 
moment, the authors succeeded in effective closing of the currrent only for fast  
pulsars when electromagnetic fields and particle number density near the light cylinder 
are high enough, so that pair creation process takes place there. In other words, at the 
moment it is impossible to close electric current without additional pair creation region 
close to the $Y$-point. On the contrary, A.Chen \& A.Beloborodov and A.Philippov \& A.Spitkovsky have
found the solutions close to the disk-dome structure in the case when outer magnetosphere 
pair production was suppressed.

\section{CONCLUSION}

Thus, now the theory of the pulsar magnetosphere and pulsar wind is a rapidly evolving field 
accumulating in the last years tens of researchers. Real progress was achieved including some 
quantitative predictions (e.g., the expressions for energy losses) which, as one might expect, 
can be directly checked. Unfortunately, pulsars give us no pure experiment, and the quantitative 
predictions are usually hidden in observations. E.g., up to now we have not identified the 
direction of the inclination angle evolution ${\dot \chi}$. Another observable parameter, 
i.e., so-called 'braking index' \mbox{$n_{\rm br} = {\ddot \Omega}\Omega/{\dot \Omega}^2$} 
also cannot be directly used for the analysis of the pulsar evolution as there are additional 
fluctuations of the value ${\ddot \Omega}$ (most likely, due to precession of neutron star, 
which has non-spherical shape) on the time scale much smaller than the dynamical life time 
$\tau_{\rm D} \sim P/{\dot P}$. For this reason, for most pulsars observations give unrealistic 
values $n_{\rm br} = \pm (10^{4}$--$10^{5})$. As a result, only for a few fast young pulsars 
their braking indexes can be used for the analysis of their energy losses. 

\begin{table}[h]
\caption{Braking indexes $n_{\rm br} = {\ddot \Omega}\Omega/{\dot \Omega}^2$ 
for fast radio pulsars (from Archibald et al, ApJ, 819, L16, 2016)}
\tabcolsep7pt\begin{tabular}{ccccccc}
\hline
 J1734+3333 & B0833-45 & J1833-1034 & B0540-69 & B0531+21  & B1509-59  & J1640-4631   \\
\hline
 0.9(2)  & 1.4(2)   & 1.857(1)  & 2.14(1)   & 2.51(1) & 2.839(1)  & 3.15(3)  \\
\hline
\end{tabular}
\label{table}
\end{table}

As shown in Table~\ref{table}, for most young pulsars observations give $n_{\rm br} \sim 3$,
which in zeroth approximation does not contradict theoretical predictions 
(below I do not include possible evolution of the magnetic field into consideration)
\begin{equation}
n_{\rm br}^{({\rm V)}} = 3 + 2 \cot^2\chi, \qquad 
n_{\rm br}^{({\rm BGI})} = 1.93 + 1.5 \, \tan^2\chi, \qquad
n_{\rm br}^{({\rm MHD})} = 
3 + \frac{\sin^2\chi \, \cos^2\chi}{(1 + \sin^2\chi)^2}.
\label{Eqn11} 
\end{equation}
On the other hand, in order to extract the evolution law from observations it is necessary to 
determine braking index $n_{\rm br}$ up to the second digit simultaneously avoiding any additional 
disturbances. That is impossible at the moment. Indeed, most of young pulsars have $n_{\rm br} < 3$,
in good agreement with BGI results. On the other hand, 'universal MHD' model predicts  
$3 < n_{\rm br} < 3.25$. But it only implies that additional precession is to be included 
into consideration to match theoretical predictions with observations. Long-term observations 
are necessary to clarify this point. Until then BGI model (with low-$\sigma$ pulsar wind and
the absence of striped current sheet) remains afloat.

\section{ACKNOWLEDGMENTS}
I thank A.Beloborodov, J.Petri and A.Philippov for very instructive discussions. This research was supported 
by Basic Research Program P-7 of the Presidium of the Russian Academy of Sciences.



\end{document}